\documentclass[twocolumn,preprintnumbers, amssymb,amsmath,aps,floatfix,prl,nofootinbib,superscriptaddress,showpacs]{revtex4-1}
\pdfoutput=1

\usepackage{epsfig}
\usepackage{bm}
\usepackage{amssymb}
\usepackage{amsmath}
\usepackage{color}
\usepackage{subfigure}
\usepackage[
            colorlinks,
            linkcolor=blue,
            anchorcolor=black,
            citecolor=blue
           ]
           {hyperref}

\newcommand{\pT} {\mbox{$p_{\mathrm{T}}$}}
\newcommand{\snn} {\mbox{$\sqrt{s_{NN}}$}}
\newcommand{\Dphi}{\mbox{$\Delta \phi$}}
\newcommand{\lr}[1]{\left\langle #1\right\rangle}
\newcommand{\qT} {q_{\mathrm{T}}}
\newcommand{\ncoll} {N_{\mathrm{coll}}}

\begin{document}

\title{
Medium-Induced Transverse Momentum Broadening via Forward Dijet Correlations}

\author{Jiangyong Jia}
\affiliation{Chemistry Department, Stony Brook University, Stony Brook, NY 11794, USA}
\affiliation{Physics Department, Brookhaven National Laboratory, Upton, NY 11796, USA}

\author{Shu-Yi Wei}

\affiliation{CPHT, CNRS, Ecole Polytechnique, Institut Polytechnique de Paris, Route de Saclay, 91128 Palaiseau, France}
\affiliation{European Centre for Theoretical Studies in Nuclear Physics and Related Areas (ECT*)\\and Fondazione Bruno Kessler, Strada delle Tabarelle 286, I-38123 Villazzano (TN), Italy}

\author{Bo-Wen Xiao}
\affiliation{Key Laboratory of Quark and Lepton Physics (MOE) and Institute
of Particle Physics, Central China Normal University, Wuhan 430079, China}

\author{Feng Yuan}
\affiliation{Nuclear Science Division, Lawrence Berkeley National
Laboratory, Berkeley, CA 94720, USA}
\begin{abstract}
Dijet azimuthal angle correlation is arguably one of the most direct probes of the medium-induced broadening effects. The evidence for such broadening, however, is not yet clearly observed within the precision of current mid-rapidity measurements at RHIC and the LHC. We show that the dijet correlation in forward rapidity from the future LHC RUN3, aided by forward detector upgrades, can reveal this broadening thanks to the steeper jet spectra, suppressed vacuum radiations and lower underlying event background, with a sensitivity comparable to that of the future high-luminosity Au+Au run at RHIC. Dijet correlation measurements at the two facilities together can provide powerful constraints on the temperature dependence of medium transport properties.
\end{abstract}
\maketitle

\textit{1. Introduction} 
 Heavy-ion collisions at RHIC and LHC create tiny droplets of strongly-coupled quark gluon matter, the Quark Gluon Plasma (QGP), that behaves like a nearly inviscid fluid and is opaque to colored probes. The properties and short-range structures of the QGP can be inferred from the scattering patterns of energetic partons/jets as they traverse the medium~\cite{Busza:2018rrf}. Previous measurements have revealed the ``jet quenching'' phenomena~\cite{Gyulassy:1990ye,Wang:1991xy}: a strong in-medium modification in the yield, shape and correlation patterns for these jets~\cite{Mehtar-Tani:2013pia,Qin:2015srf,Connors:2017ptx}. Theoretical efforts in describing these results have led to the extraction of an important parameter $\hat{q}$ quantifying the transverse momentum ($\pT$) broadening of single hard parton, which also controls jet energy loss and in-medium splitting processes\cite{Baier:1996kr,Baier:1996sk,Baier:1998kq,Zakharov:1996fv,Gyulassy:1999zd,Wiedemann:2000za,Wang:2001ifa,Arnold:2002ja,CasalderreySolana:2010eh,Casalderrey-Solana:2014bpa}. 

Unlike the typical multiple scattering process in QED, the jet probe is itself an evolving multi-body system, splitting into a shower of partons as it loses virtuality. During its propagation and evolution in the QGP, the jet not only loses energy and momentum, but also accumulates $\pT$-broadening through medium-induced radiation and scattering. While measurements of leading parton energy loss provide constrains on the $\hat{q}$ within a given model~\cite{Burke:2013yra}, they are not very sensitive to the mechanisms and formalism in the calculations. On the other hand, jet $\pT$-broadening arising from overall deflection and in-medium parton shower should be directly sensitive to any microscopic structure of the QGP. One promising observable for this purpose is the dihadron or dijet azimuthal angle $\Dphi$ correlations~\cite{Mueller:2016gko, Mueller:2016xoc,Chen:2016vem,Chen:2016cof,Chen:2018fqu,Tannenbaum:2017afg,Gyulassy:2018qhr}. In $pp$ collisions, the dijet $\Dphi$ correlation can be described using the Sudakov resummation framework established in Refs.~\cite{Banfi:2008qs,Mueller:2013wwa,Sun:2014gfa,Sun:2015doa}. In A+A collisions, the $\Dphi$ correlation is expected to be further broadened by jet-medium interactions, and this broadening, if measured, can directly constrain the $\hat{q}$. Due to large vacuum Sudakov contributions, current measurements~\cite{Aad:2010bu,Chatrchyan:2011sx,Adam:2015doa,Adamczyk:2017yhe,Sirunyan:2017qhf} are statistically and systematically limited for a clear observation.

Besides the complexity associated with the jet probes, the medium is rapidly expanding and its properties are highly dynamical. The jet-medium interactions are sensitive to the full evolution of both the jet and the medium from the initial to the final state, which complicates the determination of medium properties at given temperature. Since the mediums created at RHIC and the LHC have different temperature and different space-time evolution, a combined analysis of the same observables at RHIC and the LHC provide important lever arm to disentangle the temperature dependence from dynamical evolution~\cite{Adare:2015kwa,Akiba:2015jwa}. Such exercise has been successfully carried out for the extraction of $\eta/s$ and other bulk properties based on Bayesian analysis of soft particle observables~\cite{Bernhard:2016tnd}. In the jet sector, a simultaneous comparison to the leading hadron suppression at RHIC and the LHC was shown to reduce the certainty of $\hat{q}$, and even suggests a possible non-monotonic temperature dependence~\cite{Burke:2013yra}. This RHIC-LHC complementarity was also demonstrated for several full jet observables with the sPHENIX detector in the future RHIC run~\cite{Adare:2015kwa}.

Another lever arm for extracting the temperature dependence of QGP properties and structures is also provided by comparing jet measurements at mid-rapidity at RHIC with those at forward-rapidity at LHC. This is because the medium produced at forward-rapidity may have a temperature closer to the medium produced at mid-rapidity at lower $\snn$, but with a very different space-time dynamics. Furthermore, the jet spectra and their flavor composition at forward rapidity may also resemble those at mid-rapidity at lower $\snn$. Therefore, the forward-rapidity measurements provide a different setup for disentangling the temperature dependence from dynamical evolution. Another favorable factor is that the underlying event background (UE) fluctuations decreases at forward-rapidity, and the $dE_{\mathrm{T}}/d\eta$ at $|\eta|\sim4$ is about $\times2$ smaller than $\eta\sim0$ at the LHC~\cite{Chatrchyan:2012mb}. The measurements of forward jets and dijet correlations based on calorimetry have been demonstrated in $pp$ and $p$+Pb collisions~\cite{ATLAS:2014cpa,Aaboud:2019oop,Khachatryan:2016xdg,Sirunyan:2018ffo}. With the expected detector and luminosity upgrade in future HL-LHC, including for example the charged particle tracking to $|\eta|<4$ and improved granularity of the forward calorimetry in ATLAS and CMS experiments~\cite{Citron:2018lsq,Grosse-Oetringhaus:2018qih}, a detailed measurement of full jet and jet structure in the forward rapidity should be possible.

The complementarity between mid-rapidity LHC, forward-rapidity LHC, and mid-rapidity RHIC can in principle be demonstrated for all commonly-used jet observables. In this letter, we establish this complementary with the dijet $\Dphi$ correlation and show its potential to constrain the $\hat{q}$. Assuming integrated luminosity of 10~nb$^{-1}$ for 5.02 TeV Pb+Pb collisions expected for LHC-RUN3~\cite{Citron:2018lsq} and 100~nb$^{-1}$ for 0.2 TeV Au+Au collisions expected for sPHENIX at RHIC~\cite{Adare:2015kwa}, we estimate the expected statistical precision for dijet $\Dphi$ correlations in central collisions. We find that $\Dphi$ correlations are dominated by vacuum Sudakov contribution in mid-rapidity LHC, but are sensitive to medium-induced broadening at forward-rapidity LHC and mid-rapidity RHIC. We show that the forward-rapidity LHC provides a broader kinematic range for detecting the medium-induced broadening effects.

\textit{2. Dijet correlation in forward rapidity in $pp$ and $AA$ collisions} 
At the leading order, dijets produced in hadronic collisions are back-to-back in the azimuthal plane. However, around $\Dphi\approx\pi$ where the pair $\pT$ is much smaller than the individual jet $\pT$, the radiation of soft gluons play an important role, and its contributions need to be resummed to fully describe the experimental data. A Sudakov resummation formalism has been developed in the last few years up to next-to-leading logarithmic order for dijet $\Dphi$ correlation~\cite{Banfi:2008qs,Mueller:2013wwa,Sun:2014gfa,Sun:2015doa}, where the so-called non-global logarithmic contributions was found to be important~\cite{Dasgupta:2001sh,Dasgupta:2002bw,Banfi:2003jj,Chien:2019gyf}. Recently, this formalism has been extended to describe dijet correlations in heavy-ion collisions by including medium-induced gluon radiation~\cite{Mueller:2016gko, Mueller:2016xoc,Chen:2016vem,Chen:2016cof,Chen:2018fqu}, which we will briefly describe below.

The dijet cross section in the back-to-back limit in $pp$ collisions can be written as
\begin{align}
\frac{d\sigma_{pp}}{dPS} =& \sum_{abcd}\int \frac{d^2 b_\perp}{(2\pi)^2} e^{-i\vec q_\perp \cdot \vec b_\perp} x_a f_a(x_a, \mu_b) x_b f_b(x_b, \mu_b)\notag\\
&\times\frac{1}{\pi} \frac{d\sigma_{ab\to cd}}{d \hat t} \exp [-S (Q,b)] \; ,
\label{eq:cs}
\end{align}
where $dPS=dy_cd^2p_{Tc} dy_d d^2p_{Td}$ represents the final state phase space, $\mu_b=c_0/b_*$ with $c_0=2e^{-\gamma_E}$ and $\gamma_E$ the Euler constant. $x_a = \pT (e^{y_c}+e^{y_d})/\snn$, $x_b = \pT (e^{-y_c}+e^{-y_d})/\snn$, $Q^2 = x_a x_b S$ and $\vec q_\perp = \vec p_{Tc} + \vec p_{Td}$. $f_a(x_a, \mu_b)$ and $f_b(x_b, \mu_b)$ are the parton distribution functions. The CTEQ14 PDFs \cite{Dulat:2015mca} are used in the numerical evaluation. $d\sigma_{ab\to cd}/d \hat t$ is the partonic cross section at the leading order. By introducing the $b_*$-prescription~\cite{Collins:1984kg} which sets $b_*=b_\perp /\sqrt{1+b^2_\perp/b_{\textrm{max}}^2}$ with $b_{\rm max}=1.5 \textrm{GeV}^{-1}$, we separate the Sudakov form factor $S(Q,b_\perp)$ 
into perturbative and non-perturbative parts in $pp$ collisions: 
$S(Q,b_\perp)=S_{\mathrm{pert}}(Q,b_\perp)+S_{\mathrm{NP}}(Q,b_\perp)$ with the perturbative part defined as,
$S_{\textrm{pert}}(Q^2,b_\perp)=\int^{Q^2}_{\mu_b^2}\frac{d\mu^2}{\mu^2}
\left[A\ln\left(\frac{Q^2}{\mu^2}\right)+B+(D_1+D_2)\ln\frac{1}{R^2}\right]$, 
where $R$ represents the jet size. We have applied the anti-$k_t$
algorithm to define the final state jets in our calculations. 
Here the coefficients $A$, $B$, $D_1$, $D_2$ can be expanded
perturbatively in terms of powers of $\alpha_s$. At one-loop order, 
$A=C_A \frac{\alpha_s}{\pi}$,
$B=-2C_A\beta_0\frac{\alpha_s}{\pi}$ for gluon-gluon initial state,
$A=C_F \frac{\alpha_s}{\pi}$,
$B=\frac{-3C_F}{2}\frac{\alpha_s}{\pi}$ for quark-quark initial state,
and $A=\frac{(C_F+C_A) }{2}\frac{\alpha_s}{\pi}$,
$B=(\frac{-3C_F}{4}-C_A\beta_0)\frac{\alpha_s}{\pi}$ for gluon-quark initial state.
$D_i$ is $\frac{\alpha_s}{2\pi} C_F$ for quark jet, and $\frac{\alpha_s}{2\pi} C_A$ for gluon jet. For the non-perturbative part, we follow those in Ref.~\cite{Sun:2014gfa}.

In the $AA$ collisions, we need to add the \textit{medium} transverse momentum broadening contribution by replacing the vacuum Sudakov factor $S(Q,b_\perp)$ with the \textit{medium} modified one \cite{Mueller:2016gko}
\begin{align}
S_{\rm m} (Q,b_\perp) = S(Q,b_\perp) + \frac{1}{4} (\lr{\Delta \qT^2}_c + \lr{\Delta \qT^2}_d) b_\perp^2,
\end{align}
where $\lr{\Delta \qT^2}$ is the flavor-dependent averaged transverse momentum broadening square inside the transverse plane, and 
$\lr{\Delta\qT^2}_g = \frac{C_A}{C_F}\lr{\Delta\qT^2}_q $ between gluon jet and quark jet. Therefore, the broadening for quark jet $\lr{\Delta\qT^2}_q$ is the only free parameter in our numerical study, which is directly proportional to the $\hat q$. Following our previous estimation~\cite{Chen:2016vem}, we choose $\lr{\Delta\qT^2}_q =$ 5 GeV$^2$ and 10 GeV$^2$ for mid-rapidity in central Au+Au collisions at RHIC and central Pb+Pb collisions at the LHC, respectively. The corresponding values of $\hat q$ roughly match those extracted in previous studies~\cite{Burke:2013yra,Chen:2016vem}, which have sizable uncertainties. For the forward-rapidity LHC, we assume $5<\lr{\Delta\qT^2}_q<10$ GeV$^2$, and simply calculate for both 5 and 10 GeV$^2$ as boundary conditions. 

\begin{figure}[tbp]
\begin{center}
\includegraphics[width=0.3\textwidth]{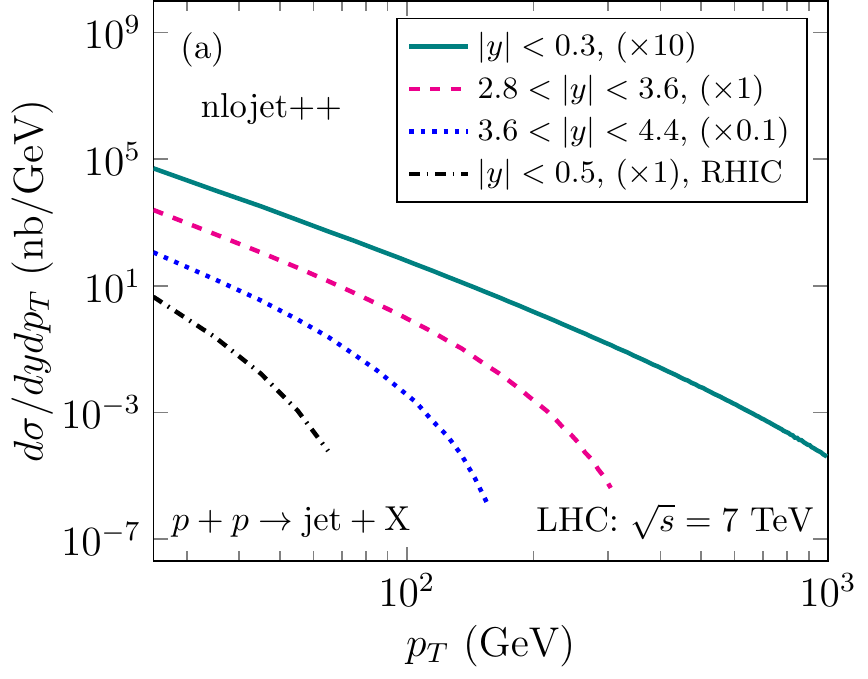}
\includegraphics[width=0.3\textwidth]{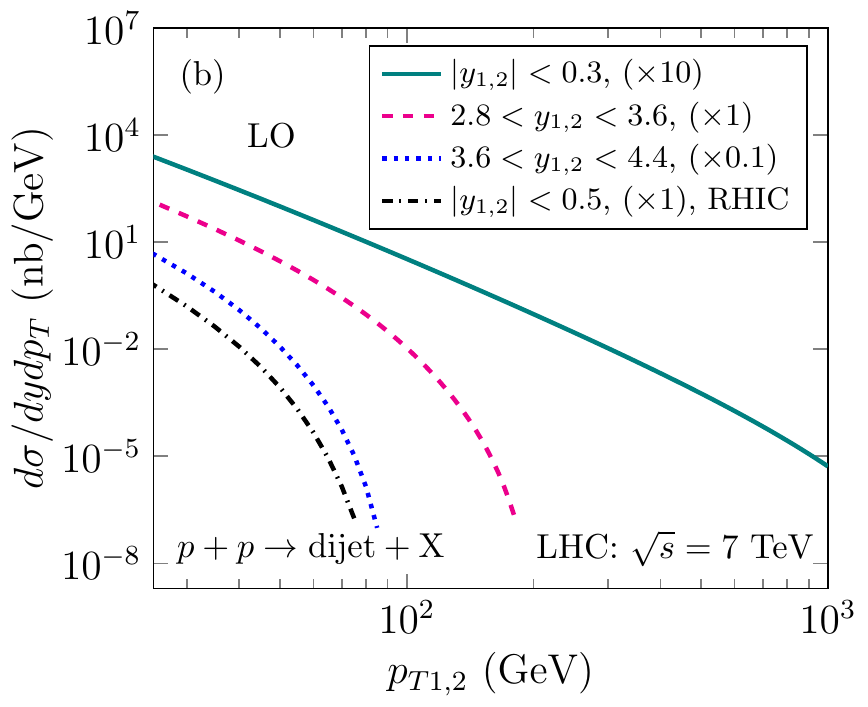}
\end{center}
\caption{Inclusive jet cross section at NLO (top panel) and dijet pair cross section at LO with both jets required in the same rapidity range (bottom panel).}
\label{fig:inclusive}
\end{figure}

Figure~\ref{fig:inclusive} shows the inclusive jet cross section in $pp$ collisions at NLO~\cite{Nagy:2001fj, Nagy:2003tz} as well as the dijet pair cross section at LO. We have checked that our NLO calculation reproduces the LHC $pp$ data.  The $\pT$ reach for inclusive jet at the LHC, even for the forward rapidity, is much larger than that at RHIC. When the away-side jet is also restricted to be in the same rapidity range, the $\pT$ reach is significantly reduced but nevertheless is still larger than that at RHIC. In principle, one could also consider cases where only one jet is in the forward rapidity, and the other jet is in the central or the backward rapidity. Such forward-central and forward-backward dijet $\Dphi$ correlation have greater pair-$\pT$ reach than forward-forward case considered here, but one need to assign two jets different $\lr{\Delta\qT^2}$ values matching their local $dN/d\eta$ values. For our first exploratory study, we choose to focus on dijet correlation with both jets restricted in the rapidity range with $3<|y|<4$, thus we can assign the same $\lr{\Delta\qT^2}$ to both jets. These forward dijets probe a QGP medium with a temperature comparable to that at RHIC but at larger $Q^2$ values.

For a quantitative estimation of the sensitivity, we assume an integrated luminosity of 10~nb$^{-1}$ for 5.02 TeV Pb+Pb expected for LHC-RUN3~\cite{Citron:2018lsq} and 100~nb$^{-1}$ for 0.2 TeV Au+Au expected for the future sPHENIX experiment at RHIC~\cite{Adare:2015kwa}. The $pp$-equivalent luminosity for the 0--10\% most central A+A is estimated as $L_{\rm pp}^{\rm AA} = f \times A^2 L_{\rm AA}$ nb$^{-1}$, where $f=R_{\rm AA}^{\rm cent} \ncoll^{\rm cent}\sigma_{\rm AA}^{\rm cent}/ (\ncoll^{\rm MB}\sigma_{\rm AA}^{\textrm{MB}})  \approx 0.2$ accounting for fractional partonic cross-section relative to minimum bias events (MB) and suppression of leading jet $R_{\rm AA}^{\rm cent}\approx 0.5$. This gives $L_{\rm pp}^{\rm PbPb}=87$~pb$^{-1}$ and $L_{\rm pp}^{\rm AuAu}=776$~pb$^{-1}$, respectively. The luminosity for $pp$ reference data is assumed to match that for the MB A+A via $L_{\mathrm{pp}} = A^2L_{\mathrm{AA}}$. From these luminosity numbers, we could estimate the number of dijet pairs in any $\Dphi$ range and the associated statistical uncertainty. For our first exploratory study, we haven't considered systematic uncertainties associated with UE background subtractions, detector resolution and harmonic flow. However, by choosing jets with small radius (e.g. $R=0.2$) and with the UE level expected in central collisions at RHIC~\cite{Adare:2015kwa}, these systematic uncertainties were shown to be reduced to a reasonable level for dijets down to 30 GeV.

\begin{figure*}[htb]
\begin{center}
\includegraphics[width=0.8\textwidth]{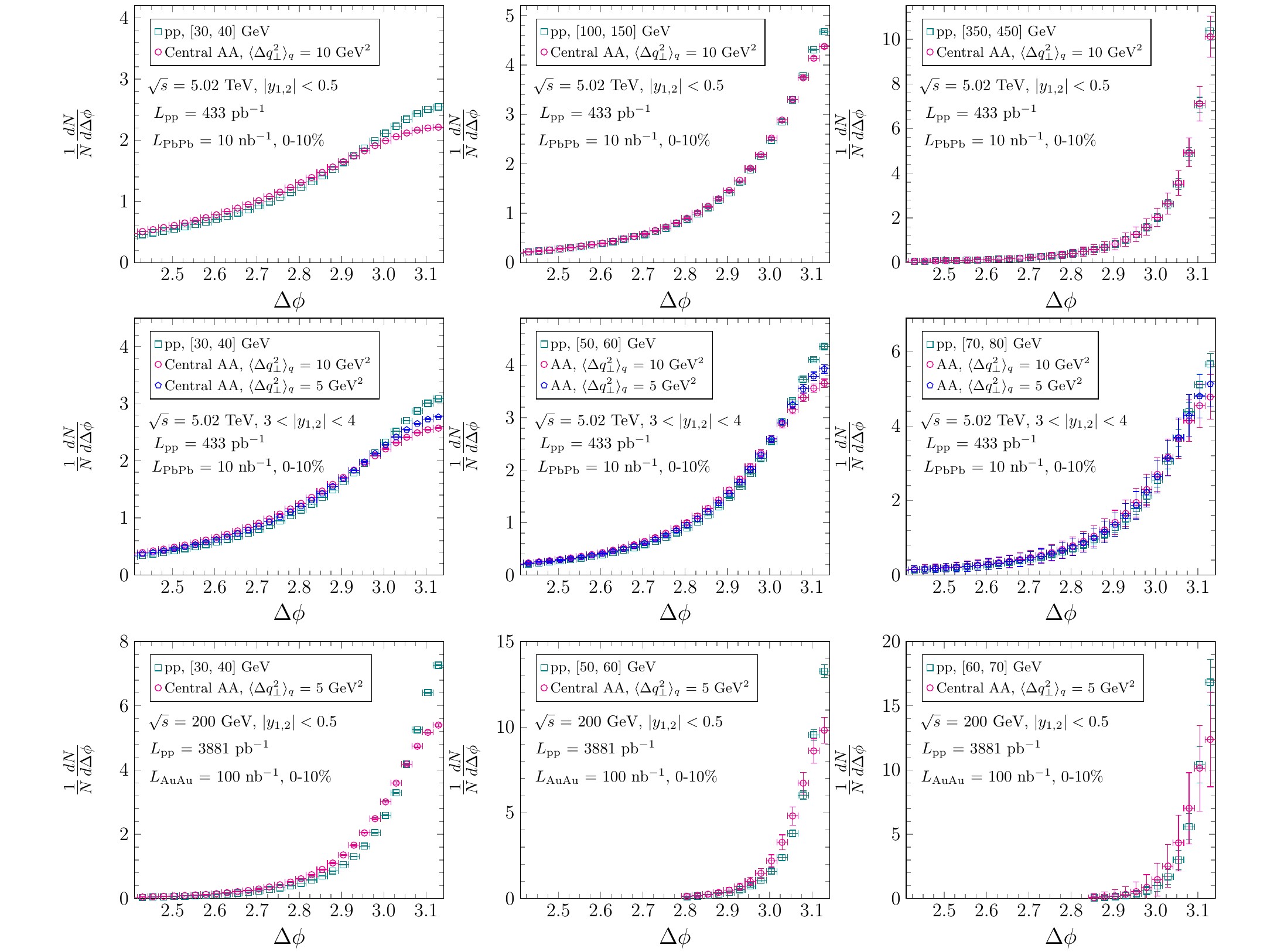}
\end{center}
\caption{Prediction for dijet $\Dphi$ distributions for the expected luminosity in the mid-rapidity at the LHC (top row),  forward rapidity at the LHC (middle row) and the mid-rapidity at RHIC (bottom row). The error bars indicate the expected statistical uncertainties.}
\label{fig:dphi}
\end{figure*}

The top-row of Fig.~\ref{fig:dphi} shows the expected dijet $\Dphi$ distributions in the mid-rapidity LHC for several $\pT$ ranges.  The corresponding dijet cross-section is large (see Fig.1), but the medium-induced broadening, reflected by the difference between Pb+Pb and $pp$, is only visible at $\pT<70$ GeV, where the full jet reconstruction is challenging due to the large UE fluctuations. At $\pT>100$ GeV where the full jet reconstruction is possible, the vacuum Sudakov factor dominates over the medium-induced broadening. This conclusion also agrees with Ref.~\cite{Mueller:2016gko}. 

The middle-row of Fig.~\ref{fig:dphi} shows the forward dijet $\Dphi$ distributions in $pp$ and central Pb+Pb for the expected luminosity at LHC-RUN3. The medium-induced broadening is more pronounced than for mid-rapidity dijet at the same $\pT$. As the UE in $3<|y|<4$ is about factor of 1.5--2 smaller than mid-rapidity~\cite{Chatrchyan:2012mb}, the forward dijets could be reconstructed at lower $\pT$ of 30--40 GeV. At higher $\pT$, the vacuum Sudakov factor is larger and difference between $pp$ and Pb+Pb is reduced. Nevertheless, the statistical precision is good enough for a possible observation up to $\pT\sim80$ GeV, especially for the $\lr{\Delta\qT^2}_q=10$ GeV$^2$ case.

We observe that the $\Dphi$ distributions in the mid-rapidity are flatter than those in the forward rapidity at the same $\pT$. This is mainly due to two reasons: 1) the Sudakov factor for the $gg$ channel that dominates in the mid-rapidity, is much larger than that for the $qg$ channel that dominates in the forward rapidity, 2) it is a common practice to set the factorization scale $\mu_f=\mu_b$ in the Sudakov resummation framework to simplify the formula, which, we find numerically, makes parton shower stronger in the middle rapidity than in the forward rapidity for $gg\to gg$ channel at low $p_T$. 

As proposed in Ref.~\cite{Mueller:2016gko},  dijet correlation at RHIC,  despite its lower $\pT$ reach,  are very sensitive to the medium-induced broadening effects due to smaller vacuum radiations. Our results for $0-10\%$ central Au+Au collisions for the expected luminosity are shown in the bottom row of Fig.~\ref{fig:dphi}.  The much smaller dijet cross section at RHIC compared to forward LHC is largely compensated by the $\times10$ larger A+A luminosity, and the medium-induced broadening is visible up to $50<\pT<60$ GeV range within the expected statistical uncertainties.

\begin{figure}[h!]
\includegraphics[width=0.45\textwidth]{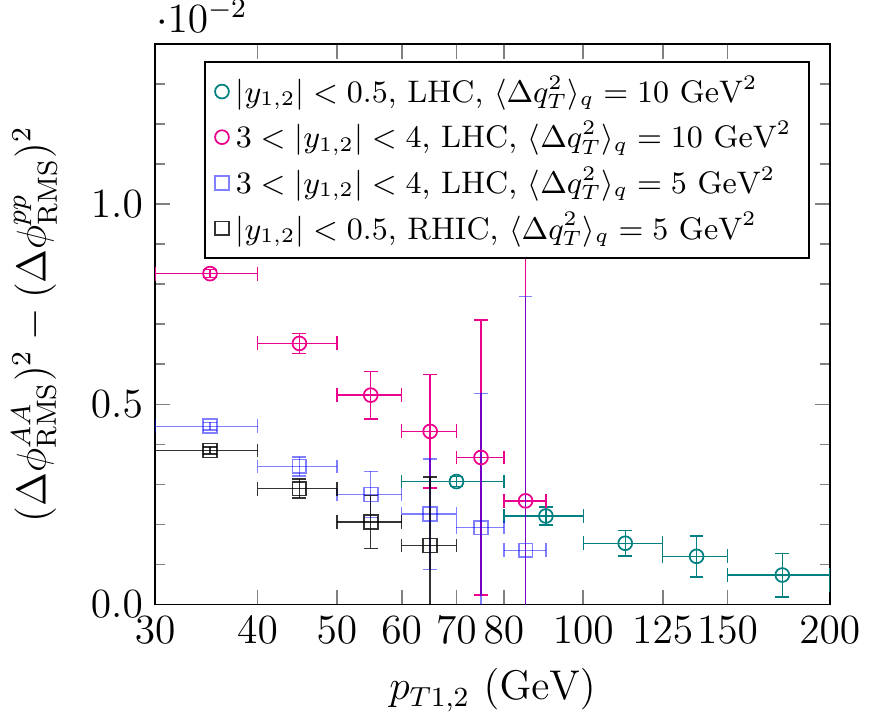}
\caption{Difference of RMS width between A+A and $pp$ collisions as a function of $\pT$. The error bars indicate the expected statistical uncertainties.}
\label{fig:rms-pT}
\end{figure}

To quantify the broadening effect, we calculate the root-mean-square (RMS) width of $\Dphi$ distribution:
\begin{align}
\Delta \phi_{\rm RMS} = \sqrt{\frac{\int d\Delta \phi (\Delta\phi - \pi)^2 \frac{d\sigma}{d\Delta\phi}}{\int d\Delta \phi  \frac{d\sigma}{d\Delta\phi}}}.
\end{align}
where the range of $\Delta\phi$ integral is chosen from 2.5 to $\pi$ at the LHC. 
From this, the difference of RMS width between A+A and $pp$ collisions is obtained to isolate the medium-induced broadening effects. Figure~\ref{fig:rms-pT} shows our results of $\Delta D\equiv(\Delta \phi_{\rm RMS}^{\textrm{AA}})^2 - (\Delta \phi_{\rm RMS}^{pp})^2$ as a function of $\pT$. In general, the $\Delta D$ is largest at low $\pT$ and decreases toward larger $\pT$. The mid-rapidity LHC results has best statistical precision, but is expected to challenging due to large UE fluctuations. The $\Delta D$ values for forward-rapidity LHC have very good statistical significance. In fact, even for $\lr{\Delta\qT^2}_q=5$ GeV$^2$, the statistical significance for forward-rapidity LHC is comparable or even slightly exceeds that for the mid-rapidity RHIC.

Figures~\ref{fig:rms-pT} shows that dijets production at RHIC is kinematically limited to about $\pT<60$ GeV (see also Fig.~\ref{fig:inclusive}), although higher-$\pT$ jets, if they were available, would in principle still be sensitive to the medium-induced broadening. In contrast, the dijet production at forward LHC covers a larger $\pT$ reach in a QGP medium that spans a range of temperature (depending on the rapidity) that could be comparable to that at RHIC. Therefore simultaneous description of dijet correlation (other jet quenching observables as well) in mid-rapidity RHIC together with the rapidity dependence at the LHC could provide powerful constraints on the jet-medium interactions.

\begin{figure}[h!]
\includegraphics[width=0.45\textwidth]{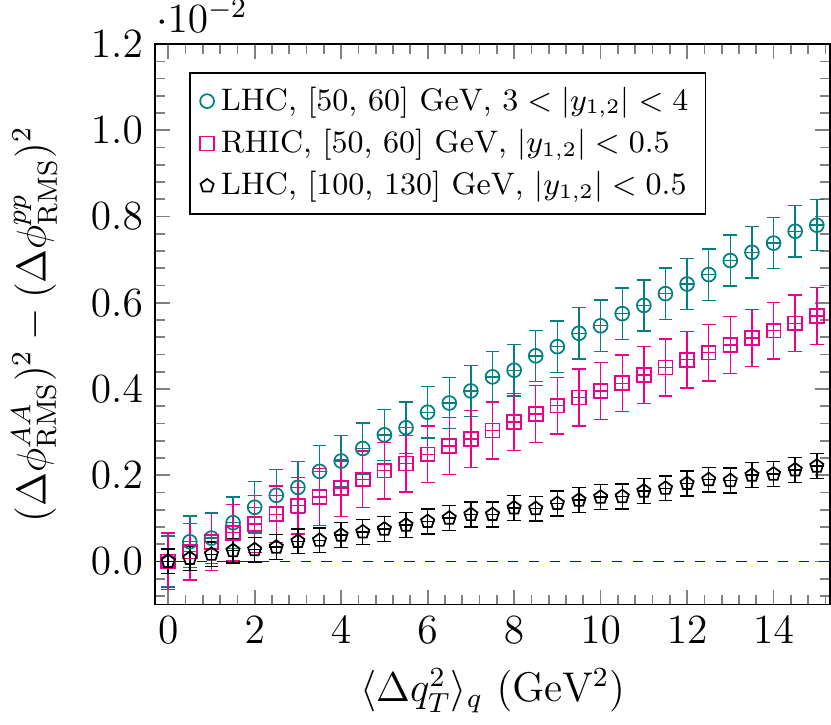}
\caption{Difference of RMS width between A+A and $pp$ collisions as a function of $\lr{\Delta \qT^2}_{q}$. The error bars indicate the expected statistical uncertainties.}
\label{fig:rms-forward}
\end{figure}

Given the present large uncertainty on the value of $\lr{\Delta\qT^2}$, we choose one representative $\pT$ range each for mid-rapidity RHIC, forward-rapidity LHC and mid-rapidity LHC, and calculate the corresponding $\Delta D$ as a function of $\lr{\Delta\qT^2}$. The results are shown in Fig.~\ref{fig:rms-forward}. The values of $\Delta D$ are comparable between RHIC and forward-rapidity LHC, and both are much larger than that for the mid-rapidity LHC. In principle, one could directly extract the value of $\lr{\Delta\qT^2}$ once $\Delta D$ is measured.

\textit{Summary} We studied the potential of forward dijet azimuthal correlation at the LHC in the search for the medium-induced $\pT$ broadening effects in heavy ion collisions. We show that the forward dijets at the LHC, enabled by future detector upgrades, are expected to have much better sensitivity compared to mid-rapidity LHC, due to the steeper jet spectra, smaller vacuum radiations and lower underlying event fluctuations. The expected sensitivity from the upcoming LHC Pb+Pb runs is comparable to that of the future Au+Au runs at RHIC, but covering a broader $\pT$ range for the dijets and a medium with different temperatures. A direct comparison of the same observable between RHIC and the LHC should provide strong constraints on the collision energy (eventually, the medium temperature) dependence of $\hat q$. 

\textit{Acknowledgment}
We thank P. Jacobs for useful comments. JJ is supported by the National Science Fundation grant number PHY-1613294 and PHY-1913138. SYW is supported by the Agence Nationale de la Recherche under the project ANR-16-CE31-0019-02. The material of this paper is based upon work partially supported by Lawrence Berkeley National Laboratory, the U.S. Department of Energy, Office of Science, Office of Nuclear Physics, under contract number DE-AC02-05CH11231, and by the Natural Science Foundation of China (NSFC) under Grant Nos.~11575070.

\end{document}